\documentclass[oneocolumn,preprintnumbers,amsmath,amssymb]{revtex4}
\pdfoutput=1

\usepackage{amsmath,latexsym,amssymb,amsfonts}
\usepackage[pdftex]{color,graphicx}
\usepackage{dcolumn}
\usepackage{bm}
\usepackage{endnotes}

\begin{document}


\title{\textbf{Why concave rather than convex inflaton potential?}}

\author{
\textsc{Pisin Chen$^{a,b}$}\footnote{{\tt pisinchen{}@{}phys.ntu.edu.tw}} and
\textsc{Dong-han Yeom$^{c,d}$}\footnote{{\tt innocent.yeom{}@{}gmail.com}}
}

\affiliation{
$^{a}$\small{Leung Center for Cosmology and Particle Astrophysics, Department of Physics, Graduate Institute of Astrophysics, National Taiwan University, Taipei 10617, Taiwan}\\
$^{b}$\small{Kavli Institute for Particle Astrophysics and Cosmology, SLAC National Accelerator Laboratory, Stanford University, Stanford, California 94305, USA}\\
$^{c}$\small{Asia Pacific Center for Theoretical Physics, Pohang 37673, Republic of Korea}\\
$^{d}$\small{Department of Physics, POSTECH, Pohang 37673, Republic of Korea}
}

\begin{abstract}
The Planck data on cosmic microwave background indicates that the Starobinsky-type model with concave inflation potential is favored over the convex-type chaotic inflation. Is there any reason for that? Here we argue that if our universe began with a Euclidean wormhole, then the Starobinsky-type inflation is probabilistically favored. It is known that for a more generic choice of parameters than that originally assumed by Hartle and Hawking, the Hartle-Hawking wave function is dominated by Euclidean wormholes, which can be interpreted as the creation of two classical universes from nothing. We show that only one end of the wormhole can be classicalized for a convex potential, while both ends can be classicalized for a concave potential. The latter is therefore more probable.
\end{abstract}

\maketitle

\newpage


\section{Introduction}

How did the universe begin? This has long been one of the most fundamental questions in physics. The Big Bang scenario, when tracing back to the Planck time, indicates that the universe should start from a regime of quantum gravity \cite{Borde:2001nh} that is describable by a wave function of the universe governed by the Wheeler-DeWitt (WDW) equation \cite{DeWitt:1967yk}. The WDW equation is a partial differential equation and hence it requires a boundary condition. This boundary condition allows one to assign the probability of the initial condition of our universe.  As is well known, to overcome some drawbacks of the Big Bang scenario, an era of inflation has been introduced \cite{Guth:1980zm}. Presumably, the boundary condition of the WDW equation would dictate the nature of the inflation.

The Planck data on cosmic microwave background (CMB) indicates that certain inflation models are more favored than some others \cite{Ade:2015lrj}. In particular, the Starobinsky-type model \cite{Starobinsky:1980te} with concave inflation potential ($V'' < 0$ when the inflation is dominant.) appears to be favored over the convex-type ($V'' > 0$) chaotic inflation \cite{Linde:1983gd}. Is there any reason for this? Here we argue that if our universe began with a Euclidean wormhole, then the Starobinsky-type inflation is probabilistically favored.

One reasonable assumption for the boundary condition of the WDW equation was suggested by Hartle and Hawking \cite{Hartle:1983ai}, where the ground state of the universe is represented by the \textit{Euclidean path integral} between two hypersurfaces. The Euclidean propagator can be described as follows:
\begin{eqnarray}
\Psi[h^{\mathrm{b}}_{\mu\nu}, \chi^{\mathrm{b}}; h^{\mathrm{a}}_{\mu\nu}, \chi^{\mathrm{a}}] = \int \mathcal{D}g\mathcal{D}\phi \;\; e^{-S_{\mathrm{E}}[g,\phi]} \simeq \sum_{\mathrm{a} \rightarrow \mathrm{b}} e^{- S_{\mathrm{E}}^{\mathrm{instanton}}},
\end{eqnarray}
where $g_{\mu\nu}$ is the metric, $\phi$ is an inflaton field, $S_{\mathrm{E}}$ is the Euclidean action, and $h^{\mathrm{a,b}}_{\mu\nu}$ and $\chi^{\mathrm{a,b}}$ are the boundary values of $g_{\mu\nu}$ and $\phi$ on the initial (say, $\mathrm{a}$) and the final (say, $\mathrm{b}$) hypersurfaces, respectively. Using the steepest-descent approximation, this path integral can be well approximated by a sum of instantons \cite{Hartle:1983ai}, where the probability of each instanton becomes $P \propto e^{-S_{\mathrm{E}}}$. This approach has been applied to different issues with success: (1) It is consistent with the WKB approximation, (2) It has good correspondences with perturbative quantum field theory in curved space (e.g., \cite{Linde:1993xx}), (3) It renders correct thermodynamic relations of black hole physics and cosmology \cite{Gibbons:1976ue}. These provide us the confidence that the eventual quantum theory of gravity should retain this notion as an effective description.

In their original proposal, Hartle and Hawking considered only compact instantons. In that case it is proper to assign the condition for only one boundary; this is the so-called \textit{no-boundary proposal}. In general, however, the path integral should have two boundaries. If the arrow of time is symmetric between positive and negative time for classical histories, then one may interpret this situation as having {\it two universes created from nothing}, where the probability is determined by the instanton that connects the two classical universes \cite{Chen:2016ask}\endnote{One can of course interpret this as a uni-directional arrow of time with a quantum big bounce. Such interpretation will not cause any difference at the homogeneous level, but will introduce a difference at the perturbative level.}. Such a process can be well described by the \textit{Euclidean wormholes}\footnote{Note that in our approach we did not integrate out the unobserved part of the universe. This is because in our setting the two universes are connected through an on-shell solution, and so there is no hidden degree of freedom to be integrated.} \cite{Hawking:1988ae} (see also \cite{RoblesPerez:2010zz}).

In this paper, we investigate Euclidean wormholes in the context of the inflationary scenario in order to answer the question on the preference of a specific shape of the inflaton potential. This paper is organized as follows. In Sec.~\ref{sec:mod}, we first describe our model and then discuss the detailed construction of the Euclidean wormholes. In Sec.~\ref{sec:pre}, we investigate various inflaton potentials and check their different characteristics in terms of Euclidean wormholes, especially in terms of the classicality. Finally, in Sec.~\ref{sec:dis}, we shortly comment on future perspectives.

\section{\label{sec:mod}Model}

Let us consider the following action
\begin{eqnarray}
S = \int \sqrt{-g}dx^{4} \left[ \frac{R}{16\pi} - \frac{1}{2} \left(\nabla \phi\right)^{2} - V(\phi) \right]
\end{eqnarray}
with the Euclidean minisuperspace metric given by
\begin{eqnarray}
ds^{2}_{\mathrm{E}} = d\tau^{2} + a^{2}(\tau) d\Omega_{3}^{2}.
\end{eqnarray}
Then the equations of motion are as follows:
\begin{eqnarray}
\label{eq:1}\dot{a}^{2} - 1 - \frac{8\pi a^{2}}{3} \left( \frac{\dot{\phi}^{2}}{2} - V \right) &=& 0,\\
\label{eq:2}\ddot{\phi} + 3 \frac{\dot{a}}{a} \dot{\phi} - V' &=& 0, \\
\label{eq:3}\frac{\ddot{a}}{a} + \frac{8\pi}{3}\left( \dot{\phi}^{2} + V \right) &=& 0.
\end{eqnarray}
If the inflaton field is constant and the potential is flat ($V(\phi) = V_{0}$), then we recover the original Hartle-Hawking no-boundary scenario. In such case the scale factor should satisfy
\begin{eqnarray}
\dot{a}^{2} + V_{\mathrm{eff}}\left(a\right) = 0,
\end{eqnarray}
where
\begin{eqnarray}
V_{\mathrm{eff}}\left(a\right) = -1 + \frac{a^{2}}{\ell^{2}}
\end{eqnarray}
and $\ell^{2} = 3/8\pi V_{0}$. Since $V_{\mathrm{eff}} < 0$ is the allowed region for the Euclidean signature, this describes a compact instanton.

\subsection{Why Euclidean wormholes?}

In general, the kinetic term can be non-vanishing and the potential non-flat. Transcribing the system from Lorentzian into Euclidean spacetime, the real-time derivative becomes purely imaginary. The above generic action then allows for a Euclidean bounce in the $a\rightarrow 0$ regime and thus induces a Euclidean wormhole. We can see this by using a simple illustration. Now let us first investigate the flat potential $V(\phi) = V_{0}$, but with a non-trivial kinetic term. The generic solution of $\phi$ in the Lorentzian signature is
\begin{eqnarray}
\frac{d\phi}{dt} = \frac{\mathcal{A}}{a^{3}},
\end{eqnarray}
where $\mathcal{A}$ is a constant. When one Wick-rotates the Lorentzian to the Euclidean time, one obtains
\begin{eqnarray}
\frac{d\phi}{d\tau} = -i \frac{\mathcal{A}}{a^{3}}.
\end{eqnarray}
By inserting this into the Euclidean Friedman equation, the result is again $\dot{a}^{2} + V_{\mathrm{eff}}\left(a\right) = 0$, where
\begin{eqnarray}
V_{\mathrm{eff}}\left(a\right) = -1 + \frac{a^{2}}{\ell^{2}} + \frac{a_{0}^{4}}{a^{4}},
\end{eqnarray}
and $a_{0}^{4} = 4\pi \mathcal{A}^{2}/3$. This simple formula shows that $V_{\mathrm{eff}}$ has two zeros $a_{\mathrm{min}} \simeq a_{0}$ and $a_{\mathrm{max}} \simeq \ell$, and the Euclidean scale factor $a$ evolves in the following sequence:
\begin{eqnarray}
a_{\mathrm{max}} \rightarrow a_{\mathrm{min}} \rightarrow a_{\mathrm{max}}.
\end{eqnarray}
In this picture, there is no way to remove the initial boundary, but the Euclidean path integral formalism still works well nonetheless.

\subsection{Inflaton potentials and initial conditions}

Now the next step is to generalize our notion to a model with a non-trivial inflaton potential.
For definiteness, we model the chaotic-type convex potential as
\begin{eqnarray}
V_{\mathrm{ch}}(\phi) = \frac{3}{8\pi \ell^{2}} \left( 1 + \frac{\mu^{2}}{2} \phi^{2} \right),
\end{eqnarray}
and the Starobinsky-type concave inflation as
\begin{eqnarray}
V_{\mathrm{st}}(\phi) = \frac{3}{8\pi \ell^{2}} \left( 1 + A \tanh^{2} \frac{\phi}{\alpha} \right), 
\end{eqnarray}
which has a flat direction in the large $\phi$ limit. Here $\ell$ is the effective Hubble radius of the de Sitter space around the local minimum. The detailed dynamics is independent of $\ell$ up to an overall rescaling of the metric; hence without the loss of generality, we choose $\ell = 2$. For numerical demonstrations, we choose $\mu = 0.1$ for $V_{\mathrm{ch}}$ while $\alpha = 1/15$ and $A = \mu^{2} \alpha^{2}/2$ for $V_{\mathrm{st}}$, so as to maintain the same effective mass around the local minimum.

\begin{figure*}
\begin{center}
\includegraphics[scale=0.23]{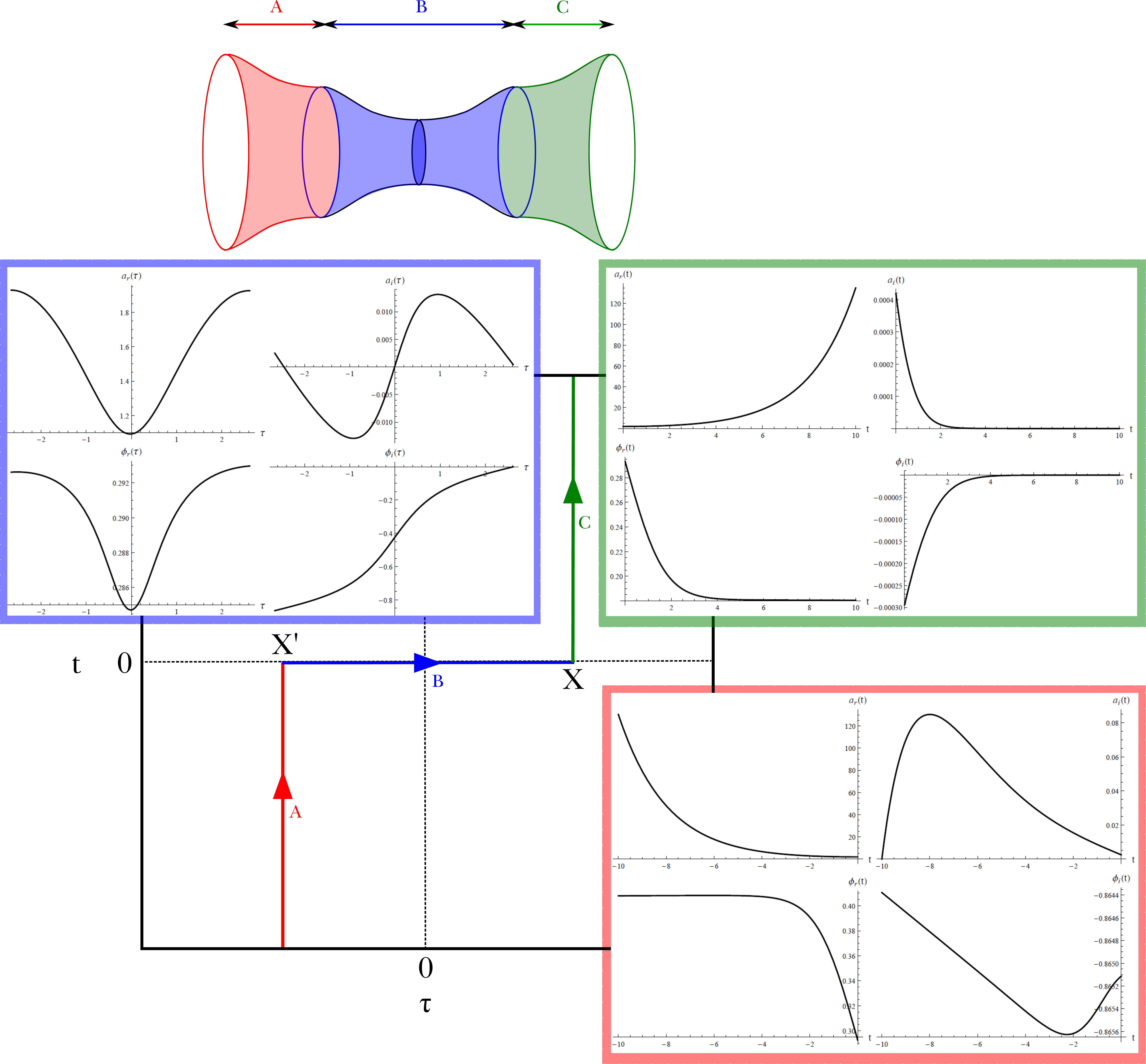}
\caption{\label{fig:FIG1}Complex time contour and numerical solution of $a_{r}$, $a_{i}$, $\phi_{r}$, and $\phi_{i}$ for $V_{\mathrm{ch}}$. The upper figure is a physical interpretation about the wormhole, where Part A (red) and C (green) are Lorentzian and Part B (blue) is Euclidean.}
\end{center}
\end{figure*}

In order to give a consistent initial condition, we impose the following ansatz:
\begin{eqnarray}
&&a_{r}(0) = a_{\mathrm{min}} \cosh \eta, \;\;\;\;\;\;\;\;\;\;\;\;\;\;\;\;\;\;\;\;\;\;\;\;\;\;\;\;\;\;\;\; a_{i}(0) = a_{\mathrm{min}} \sinh \eta,\\
&&\dot{a}_{r}(0) = \sqrt{\frac{4\pi}{3}} \frac{\mathcal{B}}{a_{\mathrm{min}}^{2}} \sqrt{\sinh{\zeta}\cosh{\zeta}}, \;\;\;\;\;\;\;\;\;\; \dot{a}_{i}(0) = \sqrt{\frac{4\pi}{3}} \frac{\mathcal{B}}{a_{\mathrm{min}}^{2}} \sqrt{\sinh{\zeta}\cosh{\zeta}},\\
&&\phi_{r}(0) = \phi_{0} \cos \theta, \;\;\;\;\;\; \phi_{i}(0) = \phi_{0} \sin \theta, \;\;\;\;\; \dot{\phi}_{r}(0) = \frac{\mathcal{B}}{a_{\mathrm{min}}^{3}} \sinh{\zeta}, \;\;\;\;\;\; \dot{\phi}_{i}(0) = \frac{\mathcal{B}}{a_{\mathrm{min}}^{3}} \cosh{\zeta},
\end{eqnarray}
where $a_{\mathrm{min}}$, $\mathcal{B}$, $\phi_{0}$, $\eta$, $\zeta$, and $\theta$ are free parameters. Here, $a_{\mathrm{min}}$ and $\eta$ should be further restricted by Eq.~(\ref{eq:1}) to satisfy the following conditions:
\begin{eqnarray}
0 &=& 1 + \frac{8\pi}{3} a_{\mathrm{min}}^{2} \left( -V_{\mathrm{re}} + 2 \cosh \eta \sinh \eta \; V_{\mathrm{im}} \right) - \frac{8\pi \mathcal{B}^{2}}{3 a_{\mathrm{min}}^{4}} \left( \frac{1}{2} + 2 \cosh \zeta \sinh \zeta \cosh \eta \sinh \eta \right),\\
a_{\mathrm{min}}^{6} &=& \frac{\mathcal{B}^{2} \cosh \eta \sinh \eta}{-2 \cosh \eta \sinh \eta \; V_{\mathrm{re}} - V_{\mathrm{im}}},
\end{eqnarray}
where $V_{\mathrm{re}}$ and $V_{\mathrm{im}}$ are the real and the imaginary part of $V(\phi)$ at $\tau = 0$, respectively.

These solutions are in general complex-valued. However, not all complex instantons are relevant for the creation of universes. After the Wick rotation to the Lorentzian time, the manifold should be smoothly connected to a classical and real-valued observer. This is called the \textit{classicality condition} \cite{Hartle:2007gi}. More formally, since we can approximately express $\Psi[q_{I}] \simeq A[q_{I}] \exp{i S[q_{I}]}$, where $q_{I}$ are canonical variables with $I=1,2,3, ...$, the classicality condition can be cast as $\left|\nabla_I A\left[q_{I}\right]\right|\ll \left|\nabla_I S\left[q_{I}\right]\right|$, for all $I$. When this classicality condition is satisfied, the corresponding history obeys the semi-classical Hamilton-Jacobi equation.

\begin{figure*}
\begin{center}
\includegraphics[scale=0.23]{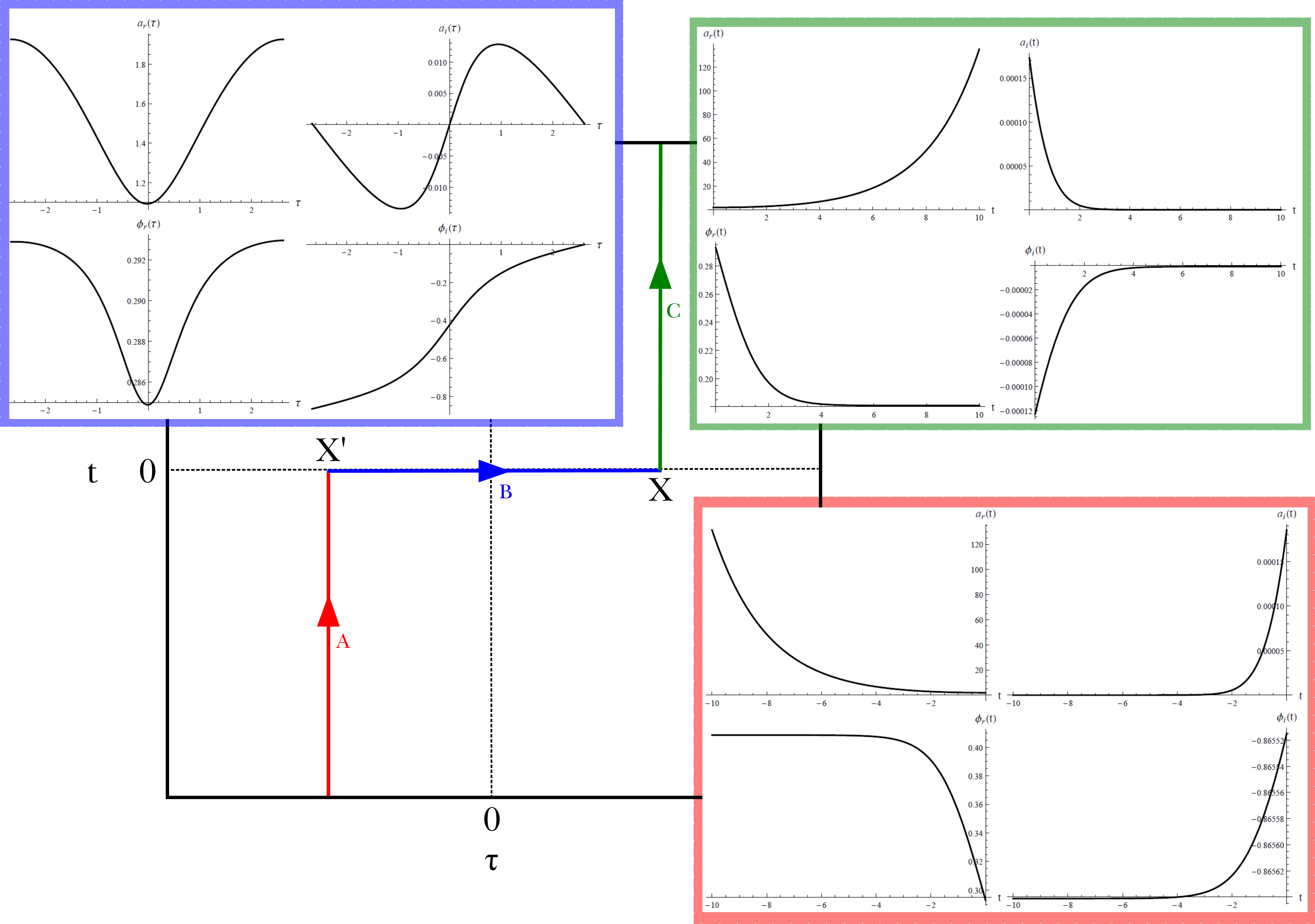}
\caption{\label{fig:FIG2}Complex time contour and numerical solution of $a_{r}$, $a_{i}$, $\phi_{r}$, and $\phi_{i}$ for $V_{\mathrm{st}}$. Now both of Part A (red) and C (green) can be classicalized.}
\end{center}
\end{figure*}

In order to satisfy the classicality condition, $\theta$ must be tuned (the same as compact instanton cases, e.g., \cite{Hwang:2011mp}). As the complex contour can be represented by $\tau + it$ (where $\tau$ and $t$ are real), one can choose a turning time $\tau = X$ where the spacetime switches from Euclidean to Lorentzian signatures. Following the Lorentzian contour $X + it$ and as $t \rightarrow \infty$, we find that the classicality condition requires $a_{i} \rightarrow 0$ and $\phi_{i} \rightarrow 0$. Consequently, a wormhole is characterized by the remaining three parameters: $\mathcal{B}$, $\zeta$, and $\phi_{0}$, where $\mathcal{B}$ is related to the size of the throat, $\zeta$ the degree of asymmetry between the two ends of the wormhole, and $\phi_{0}$ the initial field value on a given potential (see \cite{Chen:2016ask}).

These three parameters ($\mathcal{B}$, $\zeta$, and $\phi_{0}$) are related to the shape of the wormhole only and have nothing to do with classicality. Note that unless both ends of the wormhole are classicalized, its probability cannot be well defined, and hence it cannot be a legitimate solution for a consistent Euclidean quantum cosmology. To attain the classicality of not one but both ends, the only remaining knob is the shape of the inflation potential.

\section{\label{sec:pre}Preference of concave inflaton potential}

Fig.~\ref{fig:FIG1} is an example for the chaotic inflation model $V_{\mathrm{ch}}$. For the initial condition $\mathcal{B} = \sqrt{3/4\pi}$, $\phi_{0} = 0.51$, and $\zeta = 0.001$, there exists a turning time $X$ that satisfies $a_{i} \rightarrow 0$ as $t \rightarrow \infty$. By fixing $\theta \simeq 3\pi/2 + 0.5923$, we can in addition tune $\phi_{i}$ to approach zero along the $X + it$ contour. There should be another turning time $X'$ that governs the evolution of the other side of the wormhole. In this case, however, no matter how we optimize $X'$, the condition for $a_{i}$ and $\phi_{i}$ to approach zero as $t \rightarrow - \infty$ simply cannot be found. Thus this wormhole cannot be classicalized at both ends. 

\begin{figure*}
\begin{center}
\includegraphics[scale=1]{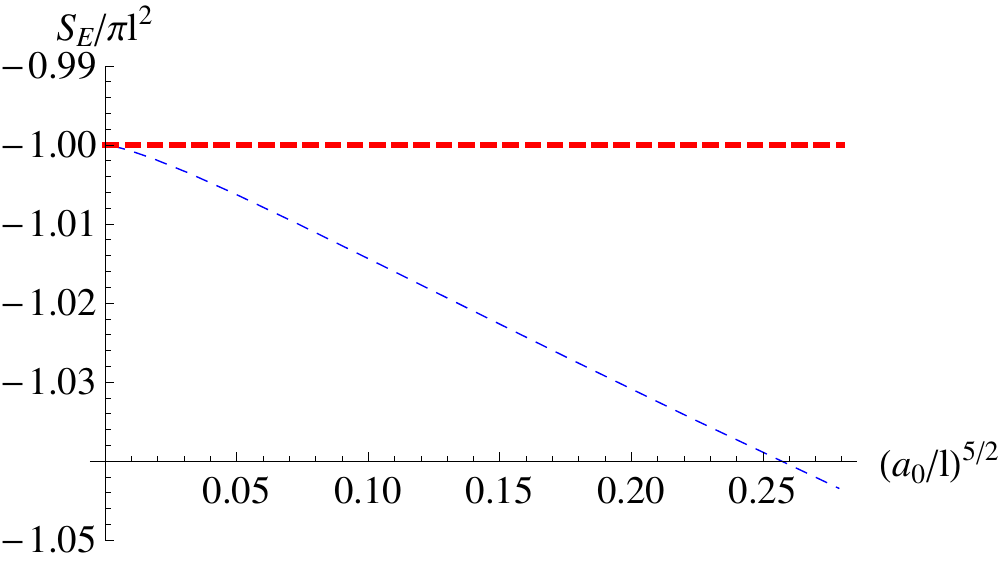}
\caption{\label{fig:action}$S_{\mathrm{E}}/\pi\ell^{2}$ as a function of $a_{0}/\ell$. The blue dashed curve is that of Euclidean wormholes, while the red dashed curve is that of compact instantons. This shows that $S_{\mathrm{E}}/\pi\ell^{2} \simeq -1 - 0.16 (a_{0}/\ell)^{5/2}$.}
\end{center}
\end{figure*}

The situation is very different for the Starobinsky-type inflation. Fig.~\ref{fig:FIG2} shows the evolution of the Starobinsky-type inflation model $V_{\mathrm{st}}$. The classicality of Part C is again obtained by tuning $\theta \simeq 3\pi/2 + 0.5927$. Here a major difference occurs in Part A. Since there is a \textit{flat direction} in the potential, $\phi_{i} \rightarrow \mathrm{constant}$ as $t \rightarrow -\infty$. As a result, the kinetic term of the scalar field vanishes and there is no contribution from the imaginary part of the scalar field. The metric therefore approaches a real-valued function, i.e., $a_{i} \rightarrow 0$. Because of this, a solution for $X'$ can be found that satisfies $a_{i} \rightarrow 0$ and $\phi_{i} \rightarrow \mathrm{constant}$ as $t \rightarrow - \infty$.

In both cases, due to the lack of free parameters for fine-tuning, there is no trivial way to make both $a_{i}$ and $\phi_{i}$ zero on both sides of the wormhole. The next option is then to make $a_{i}$ and $\phi_{i}$ vanish on one side only and make $\phi_{i}$ approach a constant on the other side, which requires that the potential should have a flat direction at least on one side. This is the reason why the Starobinsky-type model conforms with classicalized Euclidean wormholes. 

Accordingly, for a convex potential where the Euclidean wormhole cannot be classicalized, the only contribution to the probability comes from the compact instantons, which, in the slow-roll limit $V(\phi) \simeq V_{0}$, is approximately $S_{\mathrm{E}} \simeq - \pi \ell^{2}$ \cite{Hartle:2007gi,Hwang:2011mp}. On the other hand, both compact and wormhole instantons are classicalizable for a concave potential and should therefore both contribute to the probability, with the latter being
\begin{eqnarray}\label{eq:int}
S_{\mathrm{E}} = - 3\pi \int_{a_{\mathrm{min}}}^{a_{\mathrm{max}}} da \frac{a \left( 1 - a^{2}/\ell^{2} \right)}{\sqrt{V_{\mathrm{eff}}(a)}} \simeq - \pi \ell^{2} \left[ 1 + 0.16 \left( \frac{a_{0}}{\ell} \right)^{5/2} \right] \label{eq:up},
\end{eqnarray}
where $a_{0}=(4\pi \mathcal{B}^{2}/3)^{1/4}$ determines the size of the wormhole's throat (the $\zeta = 0$ limit of \cite{Chen:2016ask}). Fig.~\ref{fig:action} is a plot of the Euclidean action, and therefore the negative of the logarithm of the probability, of the compact and the wormhole instantons as a function of $a_0/\ell$. The ratio of the tunneling rates between wormholes and compact instantons is therefore $e^{+ c \pi \ell^{2}}$, where $c$ is a number depending on $(a_{0}/\ell)^{5/2} \lesssim \mathcal{O}(1)$. Since $\ell \gg 1$ for sub-Planckian inflation models, we conclude that inflation with concave potentials are exponentially more probable than that with convex potentials.

\section{\label{sec:dis}Discussion}

In this paper, we investigated Euclidean wormholes with a non-trivial inflaton potential. We showed that in terms of probability, the Euclidean path-integral is dominated by Euclidean wormholes, and only the concave potential explains the classicality of Euclidean wormholes. This helps to explain, in our view, why our universe prefers the Starobinsky-like model rather than the convex-type chaotic inflation model.

It should be mentioned that there exist other attempts to explain the origin of the concave inflation potential.  For example, in \cite{Hertog:2013mra}, it was reported that the Starobinsky-like concave potential is preferred if a volume-weighted term is added to the measure. Note that the same principle can be applied not only to compact instantons but also to Euclidean wormholes; hence, this proposal may support our result as well. We must caution, however, that the justification of such a volume-weighted term is theoretically subtle \cite{Hwang:2013nja}.

This is of course not the end of the story. One needs to further investigate whether this Euclidean wormhole methodology is compatible with other aspects of inflation. It will also be interesting to explore the relation between the probability distribution of wormholes and the detailed shapes of various inflaton potentials. Furthermore, if this Euclidean wormhole creates any bias from the Bunch-Davies state, then it may in principle be confirmed or falsified by future observations. We leave these topics for future investigations.

\section*{Acknowledgment}
DY is supported by the Korean Ministry of Education, Science and Technology, Gyeongsangbuk-do and Pohang City for Independent Junior Research Groups at the Asia Pacific Center for Theoretical Physics and the National Research Foundation of Korea (Grant No.: 2018R1D1A1B07049126). PC is supported by Taiwan National Science Council under Project No. NSC 97-2112-M-002-026-MY3 and by Taiwan National Center for Theoretical Sciences (NCTS). PC is also supported by US Department of Energy under Contract No. DE-AC03-76SF00515.

\theendnotes

\end{document}